\documentclass[fleqn,usenatbib,referee]{mnras}

\usepackage[T1]{fontenc}

\DeclareRobustCommand{\VAN}[3]{#2}
\let\VANthebibliography\thebibliography
\def\thebibliography{\DeclareRobustCommand{\VAN}[3]{##3}\VANthebibliography}

\usepackage{graphicx}   
\usepackage{amsmath}    

\newcommand{\kms}{km~s$^{-1}$} 
\newcommand{\cmmtr}{cm$^{-3}$}

\title[Complex organic molecules in the young hot core RCW\,120~S2]{Complex organic molecules in the young hot core RCW\,120~S2}

\author[M. S. Kirsanova \& A. A. Farafontova]{
Maria S. Kirsanova,$^{1,2}$\thanks{E-mail: kirsanova@inasan.ru}
Anastasiia A. Farafontova$^{1}$\\
$^{1}$Institute of Astronomy of the Russian Academy of Sciences, 119017, Pyatnitskaya str., 48, Moscow, Russia\\
$^{2}$Astro Space Centre, Lebedev Physical Institute, Russian Academy of Sciences, 117997, Profsoyuznaya str., 84/32, Moscow, Russia\\
}

\date{\today}

\pubyear{2025}
\begin{document}
\label{firstpage}
\pagerange{\pageref{firstpage}--\pageref{lastpage}}
\maketitle

\begin{abstract}
We analyse physical and chemical structure of the hot molecular core RCW\,120~S2, based on high-sensitivity (2–4~mK) APEX observations at 1.3~mm. The analysis reveals a rich molecular inventory, including complex organic molecules (COMs) such as CH$_3$OH, CH$_3$CHO, CH$_3$OCHO and CH$_3$OCH. We derive gas temperatures ($20-300$~K), H$_2$ densities ($10^{4}-10^{7}$\cmmtr), and molecular column densities. The detected emission probes a radially stratified envelope, with cooler ($\leq 60$~K) and less dense ($10^{4}-10^{5}$\cmmtr) outer layers traced by SO, SO$_2$, and c-C$_3$H$_2$, while warmer ($60-100$~K) and denser ($10^{5}-10^{7}$\cmmtr) inner regions are traced by H$_2$CO, OCS, and low-excitation CH$_3$CN. The hot gas ($\geq 100$~K) exhibits broad ($8-10$~\kms) lines from high-excitation CH$_3$OH, CH$_3$CN, HDO, and CH$_3$OCH$_3$. Relative molecular abundances of COMs generally agree with astrochemical hot-core models, while methanol appears underabundant and CH$_3$CN overabundant compared to predictions. We attribute these discrepancies to the need for interferometric observations at intermediate spatial scales to resolve the core's true filling factor and radial gradients.
\end{abstract}




\section{Introduction}\label{sec:intro}

Massive stars with $M \geq 8\,M_\odot$ form in the dense parts of molecular clouds and affect their surroundings even at the earliest stages of their evolution. During the hot core phase, the protostar is embedded in dense molecular gas. This envelope of molecular gas is compact (\(\leq 0.1\)~pc), dense ($n_{\rm H_2} \geq 10^{6}$~\cmmtr), and has gas and dust temperatures $T \geq  100$~K \citep[e.g.][]{kurtz2000, 1999PASP..111.1049G}. This phase precedes the formation of dense and compact HII region~\citep{2007prpl.conf..165B} around a massive star.

Millimeter and submillimeter wavelength observations towards the hot cores reveal rich carbon, nitrogen, oxygen, sulfur-bearing molecular chemistry~\citep{beltrán2018complexorganicmoleculeshot}. High dust temperatures of the hot core lead to the sublimation of icy dust mantles which enriches the gas with molecular species including complex organic molecules (COMs; molecules which contain carbon and consist of six or more atoms like methanol and more complex), see e.g.~\citet{2009ARA&A..47..427H}. Chemical and physical parameters of the hot cores can be estimated using different molecular tracers. Tracers of temperature include CO and molecules in which transitions between certain levels are forbidden by selection rules, e.g. different $K$ ladders of symmeric top molecules like NH$_{3}$ or CH$_{3}$CN \citep[see e.g.][]{1984ApJ...286..232L,Araya_2005}. Different K$_{-1}$ in H$_{2}$CO probe gas temperature and density in dense and warm regions~\citep{1993ApJS...89..123M}.  CH$_{3}$OH molecule as an assymetric top rotor wich has many transitions from far-infrared to millimeter wavelengths has been treated as useful molecular tracer of density and kinetic temperature of the gas~\citep{2004A&A...422..573L, 2006ARep...50..965S, Kalenskii2016}. Sulfur-bearing molecules, e.g. H$_{2}$CS, SO, SO$_{2}$, OCS, H$_{2}$S, NS, act as tracers of chemical evolution, physical and dynamical state of the star-forming regions. Species with sulfur content, e.g. SO, SO$_{2}$, may be associated with warm and turbulent gas, while the other species, e.g. H$_{2}$CS, NS, are associated with more quiescent gas in the envelope of core~\citep{Sulfur}. Chemical composition among the samples of observed hot cores have been shown to have variations \citep[e.g.][]{2022MNRAS.511.3463Q} with possible differences in their surrounding medium, age, evolutionary stage or existense of external heating sources, meaning every hot core can be unique in a particular way. 

Knowledge of the internal density, temperature, and velocity structure of cores is important for understanding the main physical processes in cores and core evolution. Hot molecular cores have radial density profiles which are often described by power law function $n \sim r^{-q}$, where parameter q typically ranges between 1.5 (free-fall collapse) and 2.0 (static isothermal sphere), see~\citet{1977ApJ...214..488S, 1985prpl.conf...81M, Shu1987}. This dependence features a sharp peak in density at the location of the young stellar object (YSO). The temperature profile of a collapsing hot molecular core \citep[based on the][model]{1977ApJ...214..488S} follows a power law, where temperature decreases with increasing radius as $T \sim r^{-p}$, where parameter $p \approx 0.4-0.75$. The application of these models to observational data, see e.g.~\citet{2019A&A...631A.142G} allowed us to reconstruct density and temperature profiles and to determine these parameters over a wide range of radii.

One of the youngest detected hot cores is the star forming region located near the south-east border of HII region RCW\,120. \citet{Figueira_2018} resolved several compact sources toward RCW\,120~S2 with ALMA. \citet{Kirsanova_2021} found high-excitation CH$_3$CN lines and approved this source as a hot core in a very early formation phase. \citet{plakitina2024} and \citet{2025AstBu..80..348P} identified 65 lines of 35 molecules in RCW~120~S2 and demonstrated that thermal desorption is responsible for appearing of methanol in the gas phase of RCW\,120~S2. These two last studies were based on observational data with the noise level about 20-40~mK in several frequency windows from approx.~200 to 260~GHz. \citet{Farafontova_2025} analysed new data obtained in almost the same frequency windows and found 407 lines of 79 molecules due to low noise level of $2-4$~K. They confirmed existence of warm and hot has toward RCW\,120~S2 and found typical widths of spectral lines of $3-5$ and $8-10$~\kms{} for these components, respectively. \citet{kirsanova2025deut} found two lines of HDO molecules belonging to the hot gas and reported that the abundance of HDO is among the lowest reported for hot cores. In the present study we continue to analyse high-sensitivity observations presented by \citet{Farafontova_2025} and constrain the physical and chemical structure of the envelope of RCW~120~S2.





\section{Observations and Methods}\label{sec:obsand methods}

We refer to \citet{Farafontova_2025} and \citet{kirsanova2025deut} for all the details related to the performed observations and their quality.

We used two approaches to estimate physical parameters of the hot core and molecular abundances in it. For those molecules, where collisional coefficients are available, we performed non-LTE modeling with the radiative transfer code RADEX~\citep{2007A&A...468..627V}. This code solves the equations of statistical equilibrium and calculates line intensities. For the solution of the radiative transfer equation the escape probability formulation is used. In this formulation it is assumed that the solution is given for an isothermal and homogeneous medium without large-scale velocity fields. In our calculations we used the case of escape probability for a static, spherically symmetric and homogeneous medium. The RADEX output represents a value for a single cell (one-zone solution) with constant gas kinetic temperature $T_{\rm kin}$, hydrogen density $n_{\rm H_{2}}$ and column density of the molecule $N$. 

The collisional coefficients for all molecules was taken from LAMDA database~\citep[][]{2005A&A...432..369S} with an exception for CH$_{3}$CCH molecule which was taken from EMAA\footnote{https://emaa.osug.fr/} database and CH$_{3}$CN molecule which was taken from~\citet{1986ApJ...309..331G}. For the H$_{2}$CO and H$_{2}$CS molecules we used a ortho-to-para ratio of 3 for the calculation of total $N$. For every molecule we created a grid of parameters $T_{\rm kin}$, $n_{\rm H_{2}}$, $N$ with dimensions 50$\times$50$\times$50. For every grid point, i.e. particular ($T_{\rm kin}$, $n_{\rm H_{2}}$, $N$) we calculated model intensity with RADEX. The ranges of grid T$_{\rm kin}$, n$_{\rm H_{2}}$, N were 10 $\leq$ T$_{\rm kin}$ $\leq$ 500 K, 10$^{3}$ cm$^{-3}$ $\leq$ n$_{\rm H_{2}}$ $\leq$ 10$^{9}$ cm$^{-3}$, 10$^{12}$ cm$^{-2}$ $\leq$ N $\leq$ 10$^{17}$ cm$^{-2}$. We calculated the $\chi^{2}$ statistics for the difference between the simulated and observed line intensities. Then with the method of global optimisation \texttt{basinhopping} from \texttt{scipy.optimize.} we found the global minimum where the $\chi^{2}$ was the lowest. The errors were refined near the global minimum with the method of local optimisation~\texttt{scipy.optimize.minimize} with solver ~\texttt{Nelder-Mead}. With this approach we got the parameters $T_{\rm kin}$, $n_{\rm H_{2}}$ and $N$ that best fit the observed line intensities. We use filling factor $f=1$ for RADEX and discuss this value below.


All the molecules that have four or more than four transitions (to constrain parameters: $T_{\rm kin}$, $n_{\rm H_2}$, N) were taken for the calculation of physical parameters of the RCW~120~YSO~S2 with RADEX. 

For those molecules for which collisional coefficients are not known we estimated excitation temperature $T_{\rm ex}$ and column density of molecules $N$ in the LTE approximation using the rotation diagram method as we did it before~\citep[see e.g.][]{plakitina2024}. For the rotational diagram analysis we took lines that have three or more than three transitions.  

\section{Results and Discussion}\label{sec:results}

Detected lines of the observed molecules are presented in Fig.~\ref{fig:observed}. Together with Fig.~4,~5~and~6 from \citet{Farafontova_2025}, these figures show all the molecules analysed in the present study. All the detected lines were fitted by a Gaussian function whose parameters are presented in \citet{Farafontova_2025}. We summarise our results in this paper and note that number of atoms of the analysed molecules varies from 2 or 3 (e.g. SO and OCS) to 8 and 9 (CH$_3$OCHO and CH$_3$OCH$_3$). 

Lines of some molecules (e.g. CH$_3$OH, CH$_3$CN and others) appear with two different widths. Namely, high-excitation lines of CH$_3$CN with the upper level energy $E_{\rm u} \geq 195$~K, CH$_3$OH with $E_{\rm u} \geq 248.9$~K, are named below as CH$_3$CN$_H$ and CH$_3$OH$_H$ because they have typical high linewidths of 7 and 8\kms, respectively. Lines with lower values of $E_{\rm u}$, more narrow with widths about $1.5-2$ times less. Therefore, we call them CH$_3$CN$_L$ and CH$_3$OH$_L$, respectively. Similarly, we will call high and low-excitation lines of acetaldehyde (CH$_3$CHO) and methyl formate (CH$_3$OCHO) 

The best-fit parameters ($T_{\rm kin}$, $n_{\rm H_2}$, $N$) calculated with non-LTE approximation are presented in the Table~\ref{tab:radex}. We find lines of CH3OH$_H$ and CH3CN$_H$ are excited in the gas with the temperature and density higher than the lines with $\Delta v=4-5$~\kms{} and lower $E_{\rm u}$. Column densities of the high-excitation methanol is less than of the low-excitation methanol about an order of magnitude, while column densities of the high- and low-excitation CH3CN are comparable within the obtained $\pm \sigma$ intervals. Looking into the Table~\ref{tab:radex}, we find that the molecules from the top are excited also in the gas with higher density than molecules from the bottom. Therefore, the analysed molecules probe the structure of the envelope of the protostar. Namely, the denser gas is more warm than the less dense because of heating by a protostar. The observed molecular emission originates from an envelope with significant radial gradients of both temperature and density.

For those molecules which have no measured collisional coefficients, we calculate excitation temperatures ($T_{\rm ex}$) and molecular column densities from LTE modelling. The LTE-results are shown in Table~\ref{tab:rotdiag}. Lines of dimethyl ether (CH$_3$OCH$_3$) can be fitted by a straight line, therefore we estimate an excitation temperature of $95 \pm 12$~K and conclude that these molecules appear in hot gas. Lines of other COMs (CH$_3$OCHO and CH$_3$CHO) can be fitted by a straight line only in the range of $E_{\rm u} \leq 120-140$~K. Therefore, we estimate  $T_{\rm ex} = 30-60$~K and column densities $\sim 10^{13}$~cm$^{-2}$ only for the low-excitation lines. While the high-excitation lines were clearly detected, their intensities can not be fitted in LTE.

Taking into account results of the LTE modelling from Table~\ref{rotdiag}, we represent the envelope of the hot core in Fig.~\ref{fig:hot_core}. Our observational data with $\sigma=2-4$~mK allowed refining thermal structure of the envelope and going deeper to the hot gas. We find such molecules as SO, SO$_2$, SiO, c-C$_3$H$_2$, CH$_3$CHO$_L$ emitting from outer layers of the hot core envelope with $T_{\rm kin} \leq 60$~K and $n_{\rm H_2} \sim 10^4-10^5$\cmmtr{} along with CH$_3$CCH and CH$_3$OH$_L$. H$_2$CO, OCS and low-excitation CH$_3$CN$_L$, CH$_3$OCHO$_L$ are observed from the region with $T_{\rm kin} \approx 60-80$~K and higher $n_{\rm H_2} \sim 10^5-10^7$\cmmtr. The high-excitation CH$_3$CN, methanol, HDO and CH$_3$OCH$_3$ appears at the highest detected temperatures and have spectral lines twice as broad as the low-excitation lines from the outer parts of the envelope.

Using the value of $N_{\rm H_2} =3.7\times10^{22}$~cm$^{-2}$ from \citet{plakitina2024}, we estimate relative abundances of molecules in the hot core and show them in Table~\ref{tab:abund}. Abundances of high-excitation CH3OH and CH3CN tend to be higher compared to the low-excitation species. The enhancement is consistent with the expectation to find more abundant complex molecules in the hot core. We note that all these abundances are obtained for the filling factor $f=1$ because we have only single-point observations here. The beam of the APEX telescope at the observed frequencies is about $(35-40)\times10^3$~AU and it is much more than expected size of a hot core \citep[see radial temperature distributions for the low mass and high mass cores e.g. in][]{2026arXiv260212792K}. \citet{Figueira_2018} resolved RCW\,120~S2 into several smaller sub-cores using ALMA and found CH3CN emission toward the biggest one with size $8.6$~mpc or 1770~AU. Therefore, abundances of the molecules emitting from the hot gas can be ever higher up to two orders of magnitude.

Comparing abundances of complex organic molecules with results of astrochemical modelling, we note that the values for such molecules as CH$_3$CHO, CH$_3$OCHO and CH$_3$OCH$_3$ are generally in agreement with the theory for hot cores~\citep[e.~g.][]{2008ApJ...682..283G, 2013ApJ...765...60G} and even for less evolved star-formation stages \citep[][]{2025ApJ...990..163B}. Probably we mostly see these COMs not from the hot gas with $T_{\rm kin} > 100$~K but from warm with lower temperatures. At least it is seen for CH$_3$CHO and CH$_3$OCHO. 

The abundance of methanol is much less than predicted $10^{-5}$ value in the mentioned models of hot cores. While applying $f\sim 0.01$ helps to raise the observed abundance of methanol, it means we should apply this value to other molecules emitting from the hot core: HDO and CH$_3$OCH$_3$. For the latter molecule it means abundance $\sim 10^{-6}$ which is two orders of magnitudes higher than the values obtained in astrochemical models \citep{2013ApJ...765...60G}. Moreother, application of $f\sim 0.01$ to the high-excited CH$_3$CN molecules would mean abundances of $\sim 10^{-5}$ which is by $3-4$ orders higher than the value from astrochemical predictions. We believe the described problem can not be resolved without interferometric observations of RCW\,120~S2 at the intermediate scales between the present single-dish and the ALMA data.

\citet{plakitina2024} performed non-LTE modelling of the low-excitation methanol lines using energy levels and collisional coefficients from \citet{2018msa..conf..276S} based on a model by \citet{2005MNRAS.360..533C}, where methanol levels with $E_{\rm u}$ up to 2500~K were used. They obtained physical parameters and methanol abundances comparable with our present calculations where collisional coefficients were taken from \citet{2010MNRAS.406...95R}, who used only levels with $E_{\rm u} < 1000$~K. Recently, \citet{2025INASR..10...10F} proposed that the former approach is preferable for high-excitation methanol lines than the latter. We refer to the used methanol energy levels and collisional coefficients as a possible reason for low methanol abundances from the analysis of the high-excitation lines. 

We also would like to note that the absence of collisional coefficients for COMs makes non-LTE calculations impossible for them. Although the LAMBDA database contains collisional coefficients for CH$_3$OCHO from \citet{2014ApJ...783...72F}, the energy level structure in that data does not match the levels probed by the emission lines we observed.

\section{Conclusions}

We study the physical conditions and molecular abundances in the hot core RCW\,120~S2 using high-sensitivity (down to $2-4$~mK) APEX observations at 1.3~mm. The hot core demonstrates emission of complex organic molecules, including CH$_3$OH, CH$_3$CHO, CH$_3$OCHO, CH$_3$OCH, together with the N-bearing CH$_3$CN. Analysis of the spectral lines using LTE and, where available, non-LTE methods shows that the abundances of CH$_3$CHO, CH$_3$OCHO and CH$_3$OCH iare in agreement with astrochemical models of hot cores. However, we simultaneously find a low methanol abundance and high CH$_3$CN abundance. Higher-resolution interferometric observations will help to resolve the structure of the hot core envelope and likely explain the observed abundances.

\section*{Acknowledgments}
This research has made use of spectroscopic and collisional data from the EMAA database (https://emaa.osug.fr and https://dx.doi.org/10.17178/EMAA). EMAA is supported by the Observatoire des Sciences de l’Univers de Grenoble (OSUG). 

\bibliographystyle{mnras}
\bibliography{main}

\clearpage

\begin{table*}
\centering
\caption{Gas temperature and density as well as molecular column densities from non-LTE modelling. Parameters of the best-fit models are shown along with their $\pm \sigma$~intervals. $^*$Parameters of the HDO line emission were taken from \citet{kirsanova2025deut}. Molecules are sorted according to the best-fit $T_{\rm kin}$ values.}
\label{tab:radex}
\begin{tabular}{lccc}
\hline
Molecule & $T_{\rm kin}$        & lg($n_{\rm H_2}$)                      & $N$  \\
         & [K]                  & [cm$^{-3}$]                            & [cm$^{-2}$]  \\
\hline
HDO$^*$                         & $308^{+400}_{-218}$& $(1.00^{+100}_{-0.03}) \times 10^{11}$ & $6.8^{+6.8}_{-4.6}\times 10^{13}$ \\
CH$_3$OH$_H$           & $196^{+20}_{-85}$  & $(2.4^{+2.5}_{-0.5}) \times 10^{7}$ & $(7.5^{+992.0}_{-2.3})\times 10^{13}$   \\
H$_2$CS                         & $99^{+30}_{-18}$   & $(3.7^{+9.6}_{-2.8}) \times 10^{6}$     & $(2.4^{+128.0}_{-0.4})\times 10^{13}$   \\
CH$_3$CN$_H$           & $89^{+59}_{-10}$   & $(7.9^{+2.1}_{-7.9}) \times 10^{7}$ & $(1.0^{+0}_{-0.1})\times 10^{16}$  \\
H$_2$CO                         & $87^{+213}_{-77}$  & $(1.2^{+8.8}_{-1.1}) \times 10^{7}$    & $(2.3^{+98.0}_{-1.9})\times 10^{14}$\\
CH$_3$CN$_L$           & $79^{+20}_{-20}$   & $(2.3^{+7.2}_{-2.1}) \times 10^{6}$ & $(1.3^{+8.7}_{-1.1})\times 10^{15}$  \\
OCS                             & $77^{+3}_{-35}$    & $(1.4^{+100.00}_{-0.10}) \times 10^{5}$   & $(3.4^{+97.0}_{-2.7})\times 10^{14}$   \\
CH$_3$OH $\Delta v=4.5$         & $56^{+20}_{-15}$   & $(7.5^{+1.5}_{-5.5}) \times 10^{6}$   & $(6.0^{+7.0}_{-4.8})\times 10^{14}$    \\
SO$_2$                          & $50^{+0}_{-45}$    & $(1.5^{+99.0}_{-1.4}) \times 10^{6}$   & $(1.2^{+99.0}_{-0.8})\times 10^{13}$    \\
CH$_3$CCH                       & $50^{+2}_{-4}$     & $(5.4^{+1.5}_{-2.0}) \times 10^{4}$    & $(6.0^{+77}_{-4.4})\times 10^{14}$  \\
SO                              & $17^{+21}_{-6}$    & $(1.0^{+1.0}_{-0.1}) \times 10^{4}$    & $(6.9^{+3.1}_{-6.8})\times 10^{15}$ \\
\hline
\end{tabular}\label{radex}
\end{table*}

\begin{table*}
\centering
\caption{Excitation temperatures and molecular column densities from LTE modelling. Parameters of the best-fit models are shown along with their $\pm \sigma$~intervals. Molecules are sorted according to the best-fit $T_{\rm ex}$ values.}
\label{tab:rotdiag}
\begin{tabular}{lcc}
\hline
Molecule         & $T_{\rm ex}$       & $N$  \\
                 & [K]                & [cm$^{-2}$] \\
\hline
$^{13}$CH$_3$OH  & $119 \pm 25$ & $(9 \pm 3)\times 10^{13}$  \\
CH$_3$OCH$_3$    & $95 \pm 12$  & $(5 \pm 1)\times 10^{14}$  \\
CH$_3$OCHO$_L$   & $65 \pm 6$   & $(7 \pm 1)\times 10^{13}$  \\
SO$_2$           & $60 \pm 10$  & $(3 \pm 1)\times 10^{13}$  \\
CH$_3$CHO$_L$    & $30 \pm 2$   & $(3 \pm 1)\times 10^{13}$  \\
SO               & $24 \pm 1$   & $(1 \pm 1)\times 10^{14}$  \\
c-C$_3$H$_2$     & $12 \pm 8$   & $(6 \pm 4)\times 10^{12}$  \\
\hline
\end{tabular}\label{rotdiag}
\end{table*}

\begin{table*}
\centering
\caption{Abundances of the detected molecules relative to H$_2$ and under assumption of filling factor $f=1$. For SO and SO$_2$ we provide the broader range between those obtained with LTE and non-LTE methods.}
\begin{tabular}{lc}
\hline
Molecule         & Abundance\\
\hline
SO               & $(0.03-3)\times 10^{-7}$  \\
OCS              & $(0.02-2)\times 10^{-7}$  \\
SO$_2$           & $(0.01-3)\times 10^{-8}$ \\
HDO              & $(0.6-4)\times 10^{-9}$  \\
H$_2$CO          & $(0.01-3)\times 10^{-7}$  \\
H$_2$CS          & $(0.05-3)\times 10^{-8}$  \\
c-C$_3$H$_2$     & $(0.5-3)\times 10^{-10}$  \\
CH$_3$OH$_H$     & $(0.01-3)\times 10^{-7}$  \\
CH$_3$OH$_L$     & $(0.3-4)\times 10^{-8}$  \\
$^{13}$CH$_3$OH  & $(1-3)\times 10^{-9}$  \\
CH$_3$CN$_H$     &  $(2-3)\times 10^{-7}$  \\
CH$_3$CN$_L$     & $(0.05-3)\times 10^{-7}$  \\
CH$_3$CCH        & $(0.04-2)\times 10^{-7}$  \\
CH$_3$CHO$_L$    & $(0.5-1)\times 10^{-9}$  \\
CH$_3$OCHO$_L$   & $(1-3)\times 10^{-9}$  \\
CH$_3$OCH$_3$    & $(1-2)\times 10^{-8}$  \\
\hline
\end{tabular}\label{tab:abund}
\end{table*}

\clearpage

\begin{figure*}
\centering
\includegraphics[width=0.6\linewidth]{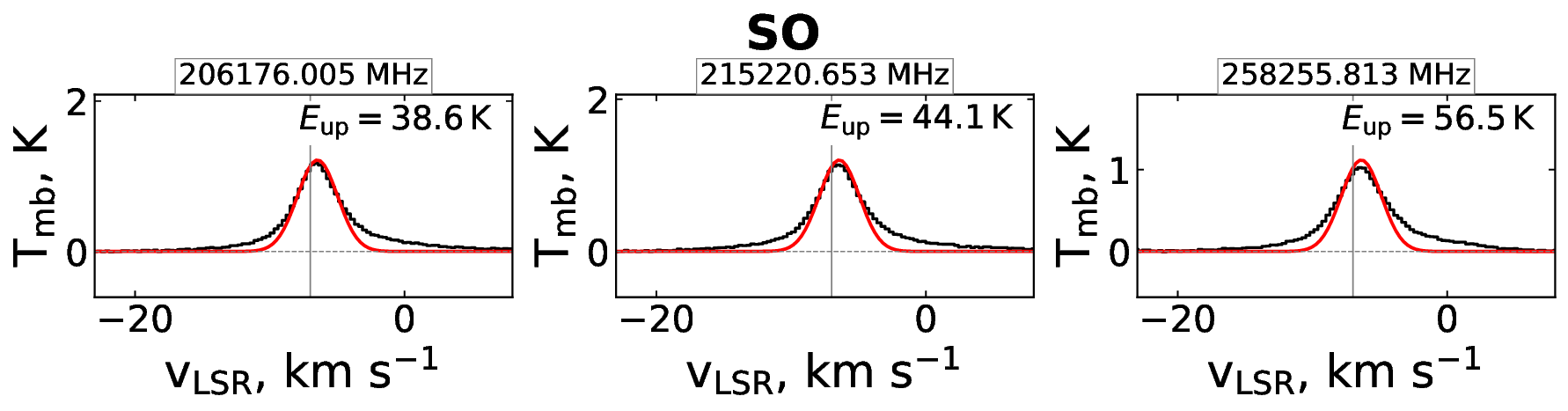}
\includegraphics[width=0.7\linewidth]{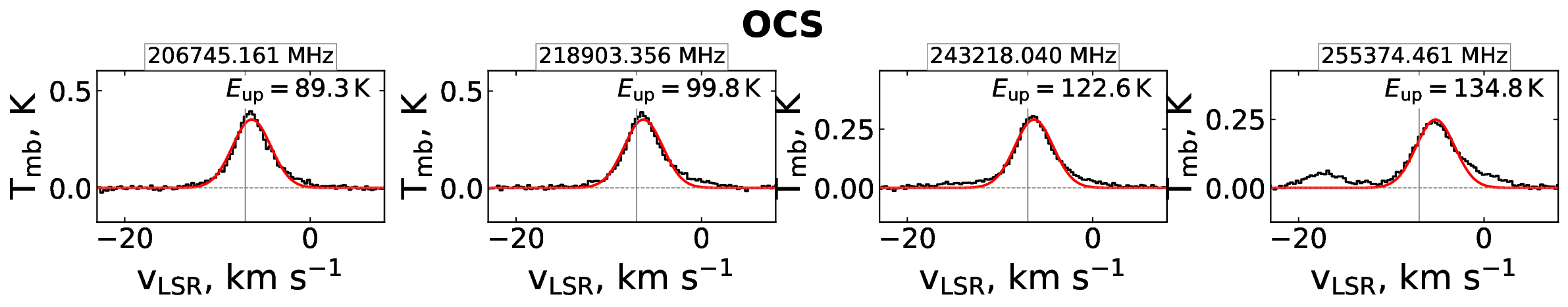}
\includegraphics[width=1.0\linewidth]{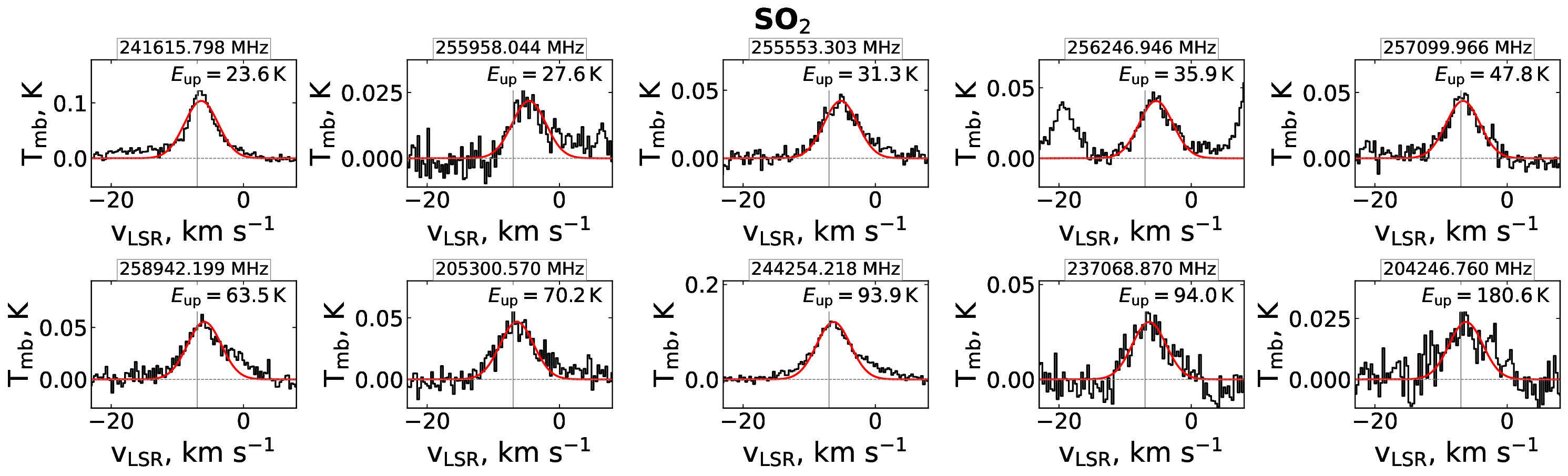}
\includegraphics[width=1.0\linewidth]{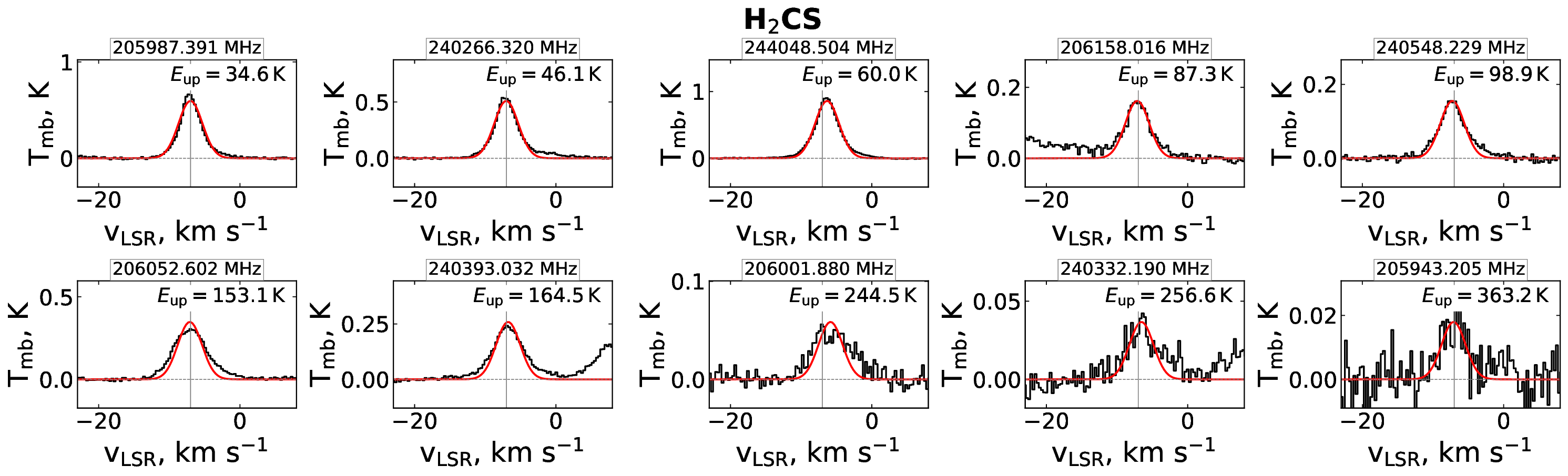}
\includegraphics[width=1.0\linewidth]{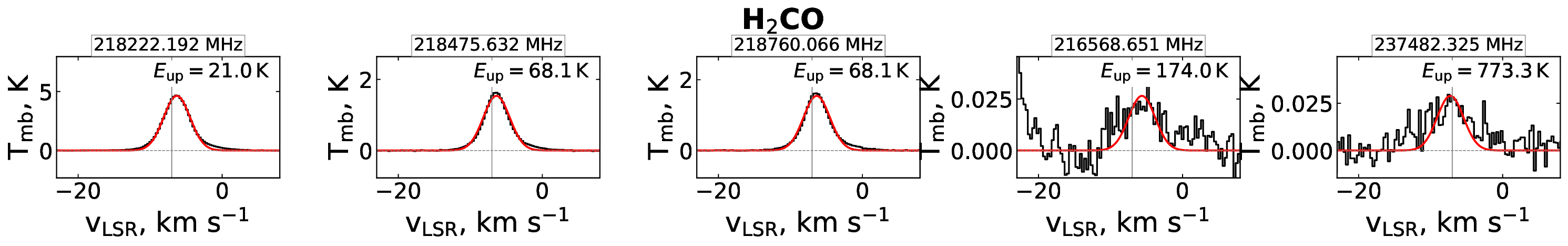}
\caption{}
\label{fig:observed}
\end{figure*}

\begin{figure*}
\centering
\includegraphics[width=1.0\linewidth]{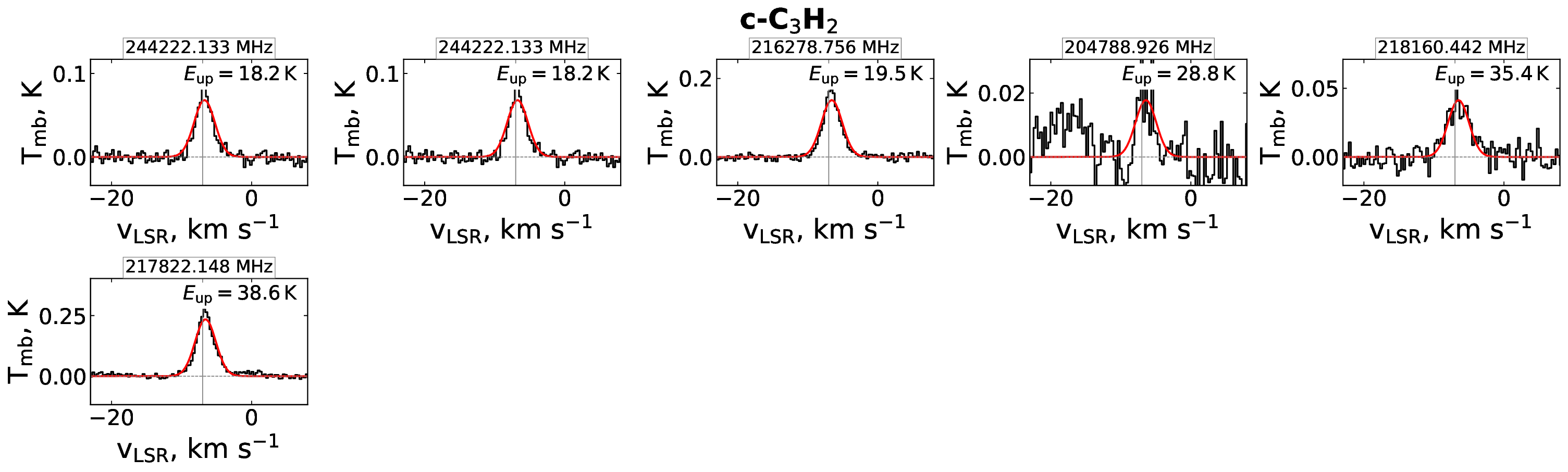}
\includegraphics[width=1.0\linewidth]{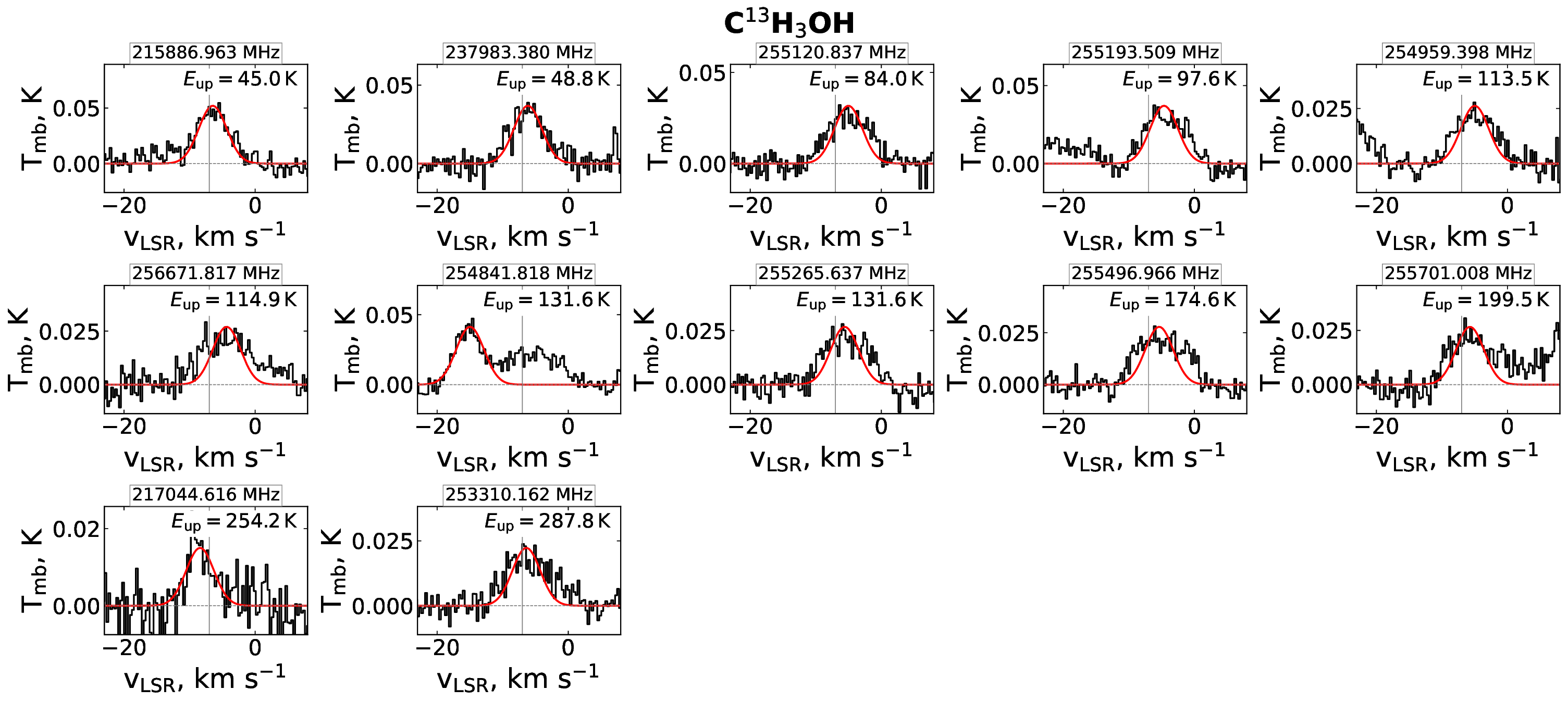}
\includegraphics[width=1.0\linewidth]{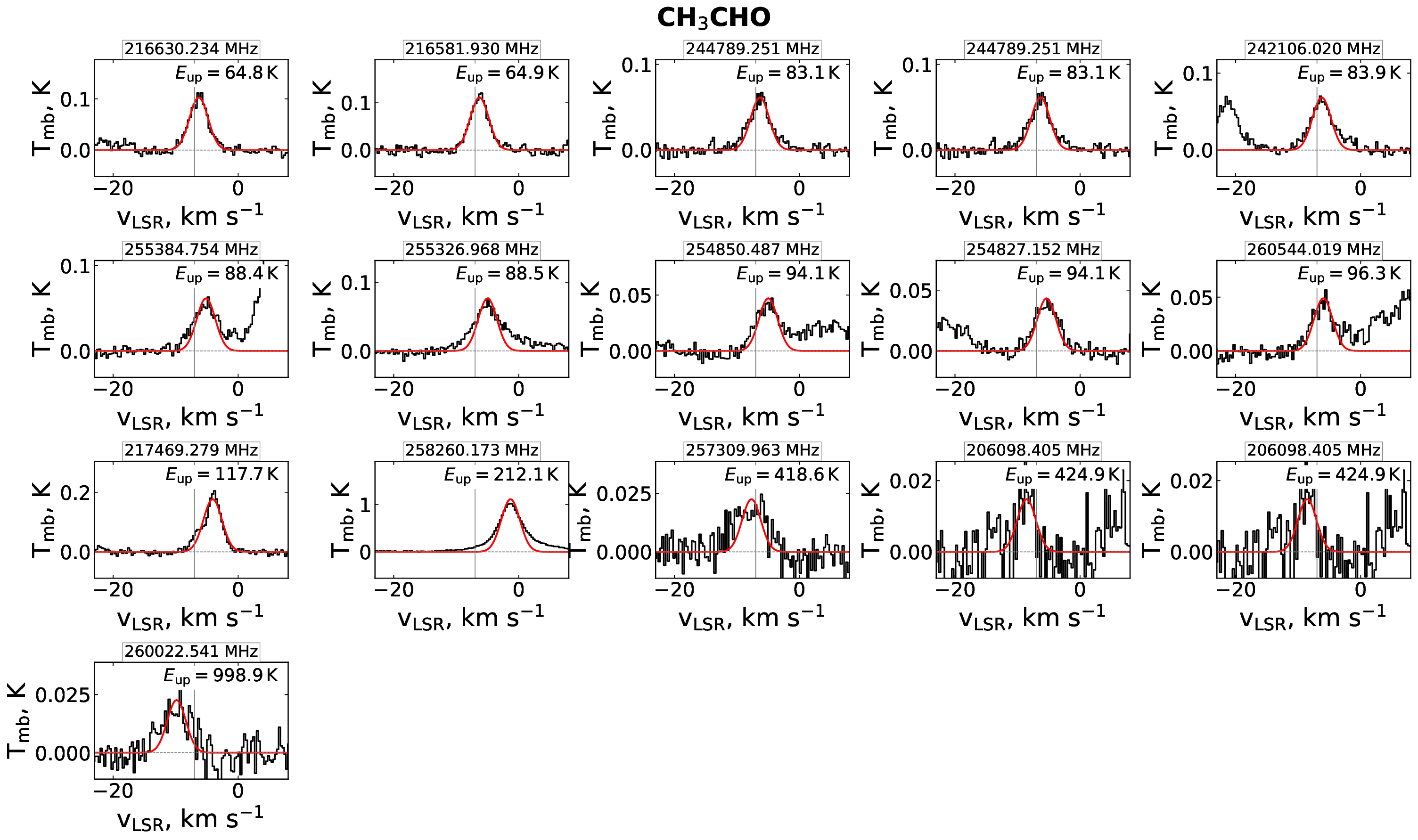}
\contcaption{Fig.~\ref{fig:observed}}
\end{figure*}

\begin{figure*}
\centering

\includegraphics[width=1.0\linewidth]{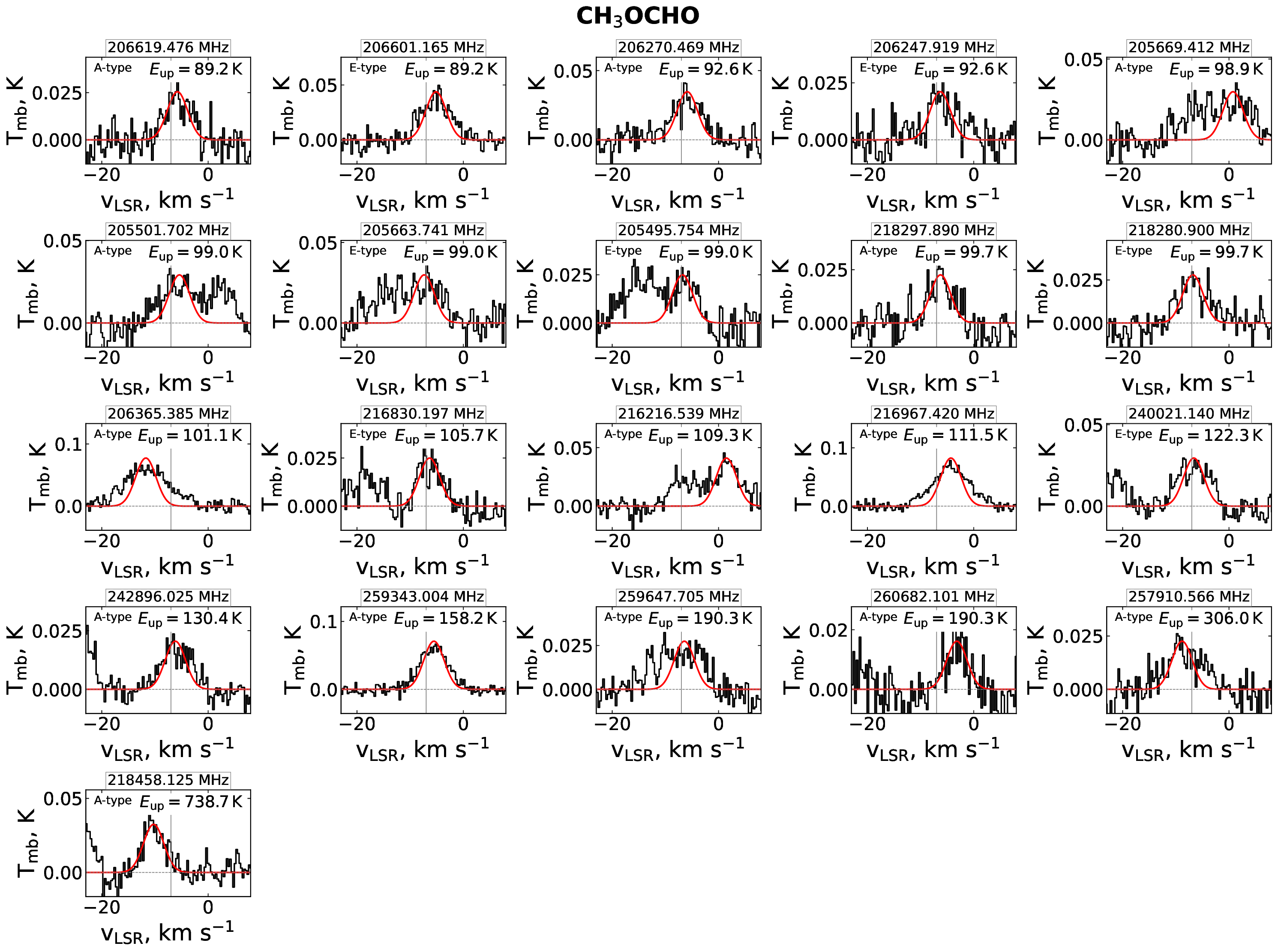}
\includegraphics[width=1.0\linewidth]{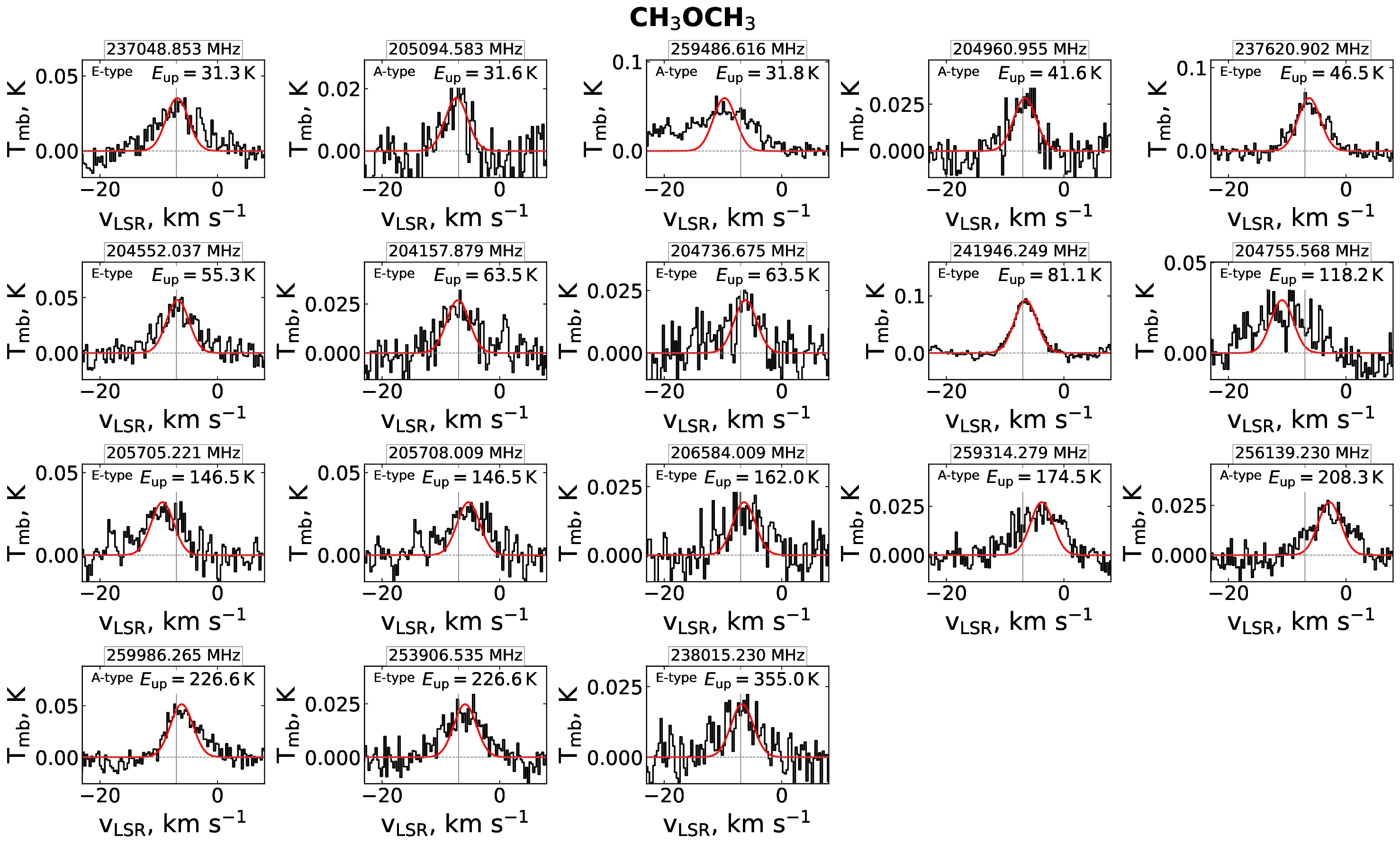}
\contcaption{Fig.~\ref{fig:observed}}
\end{figure*}


\begin{figure*}
\centering
\includegraphics[width=0.7\linewidth]{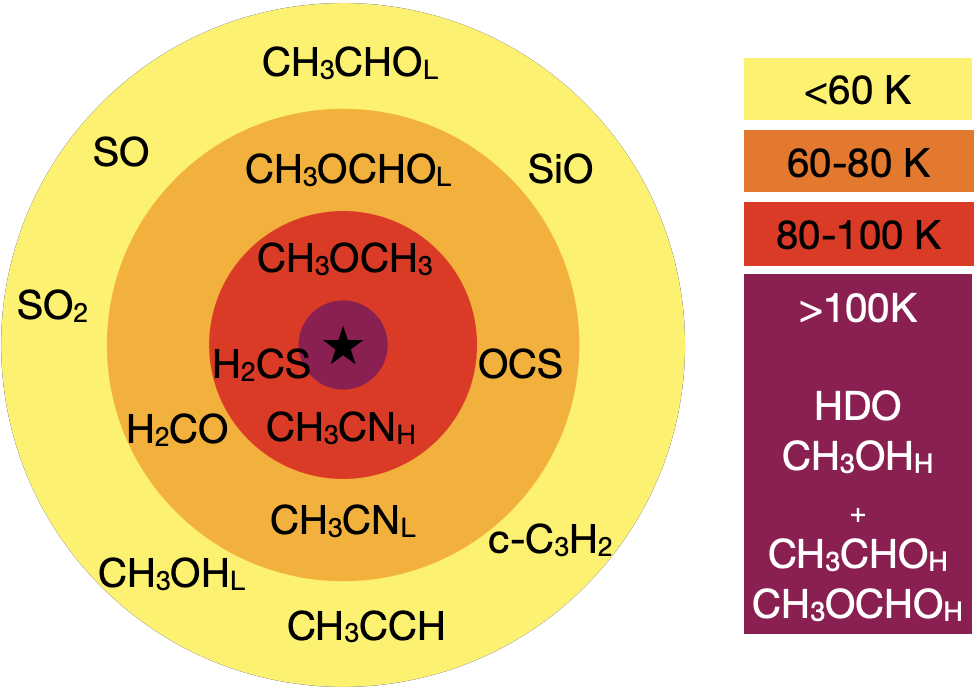}   
\caption{}
\label{fig:hot_core}
\end{figure*}

\clearpage
\section*{Captions to figures}

Figure 1. Detected molecular lines.

Figure 2. Temperature structure (not to spatial scale) of the hot core RCW\,120~S2. Left: based on data from our previous studies, Right: refined structure obtained in this study.


\end{document}